# Toward music-based stress management: Contemporary biosensing systems for affective regulation

**Natasha Yamane[1], Varun Mishra[1], and Matthew S. Goodwin[1]**
[1]Khoury College of Computer Sciences & Bouvé College of Health Sciences, Northeastern University, Boston, MA, USA

## ABSTRACT

In the last decade, researchers have increasingly explored using biosensing technologies for music-based affective regulation and stress management interventions in laboratory and real-world settings. These systems—including interactive music applications, brain-computer interfaces, and biofeedback devices—aim to provide engaging, personalized experiences that improve therapeutic outcomes. In this scoping and mapping review, we summarize and synthesize systematic reviews and empirical research on biosensing systems with potential applications in music-based affective regulation and stress management, identify gaps in the literature, and highlight promising areas for future research. We identified 28 studies involving 646 participants, with most systems utilizing prerecorded music, wearable cardiorespiratory sensors, or desktop interfaces. We categorize these systems based on their biosensing modalities, music types, computational models for affect or stress detection and music prediction, and biofeedback mechanisms. Our findings highlight the promising potential of these systems and suggest future directions, such as integrating multimodal biosensing, exploring therapeutic mechanisms of music, leveraging generative artificial intelligence for personalized music interventions, and addressing methodological, data privacy, and user control concerns.

## 1. INTRODUCTION

Stress, a ubiquitous aspect of human experience, is a departure from homeostasis provoked by a psychological, environmental, or physiological trigger.[1] Although brief episodes of stress can be beneficial, chronic exposure can elevate the risk of depression, anxiety, cardiovascular disease, and weakened immune system functioning.[2] Recognizing and managing stress and addressing its underlying risk factors (e.g., emotion dysregulation, trauma, sleep disturbance, poor diet, lack of exercise) are pivotal for promoting overall well-being and mitigating potential adverse outcomes.[3]

Thanks to advances in mobile and ubiquitous computing, stress detection and management systems have grown in number and complexity, seamlessly integrating computing capabilities into daily life.[4–6] Consumer-grade wearable biosensing technology has become increasingly prevalent in the past decade, contributing real-time, continuous, and measurable indicators of physiological arousal changes. These wearables are typically designed to interface with other devices, such as smartphones, via Bluetooth or the Internet, enabling real-time data transmission, logging, and analysis. This connectivity is critical in stress management applications as it facilitates timely feedback, personalized insights, and adaptive interventions. As a result, digital tools designed to detect and manage stress leverage biosensors to enhance their effectiveness, incorporating physiological sensing, artificial intelligence (AI), immersive experiences, or a combination thereof. This trend reflects a growing focus on precision health and individualized interventions in mental health.[4] By incorporating technologies such as electroencephalography (EEG), cardiovascular and respiration monitors, electrodermal activity (EDA) or galvanic skin response (GSR) sensors, or eye tracking devices, systems integrated with biosensors can offer real-time monitoring of physiological indicators and enable more tailored, adaptive therapeutic experiences based on users' unique biosignals.[7] Further, when developed with adaptive biofeedback capability, these systems can provide visual, auditory, and haptic feedback to users who want to manage their stress proactively.[8] Figure 1 outlines standard biosensing technologies used in affective regulation and stress management technologies.

To this end, researchers interested in biosensing-based affective regulation, stress detection, and stress management often use the circumplex model of affect[9,10] (Figure 2) to guide system design and interpret user data. This model proposes that affective states arise from two fundamental neurophysiological systems—one related to valence (a pleasure–displeasure continuum) and the other to arousal (high–low activation). By mapping individuals' responses onto a valence-arousal space, researchers can investigate the interplay between affective states (e.g., anxiety, frustration) and physiological changes in response to various stressors to explain higher-level processes such as emotional regulation and stress management.[11]

Stress management systems that do not harness biosensing technology are also available. Such tools incorporate personalized goal setting and evidence-based techniques from cognitive behavioral therapy and positive psychology, or leverage immersive platforms to reduce stress. For example, many non-biosensing mobile stress management applications integrate visual and auditory cues with cognitive restructuring or mindfulness practices to guide users through meditation exercises or identify stress-inducing thought patterns.[12] Virtual reality technologies can create immersive environments conducive to relaxation, such as natural scenes with greenery, animals, water, and rocks, among other elements.[13] Compared to biosensor-integrated systems, non-biosensing systems are typically more cost-effective as they eliminate the need for specialized sensors or wearables. However, such systems frequently lack objective outcome measurements, increase user burden, and reduce long-term adherence, thus limiting their scalability and overall use.[14] Additionally, the absence of objective outcome measurements provided by biosensors makes validating their effectiveness difficult.

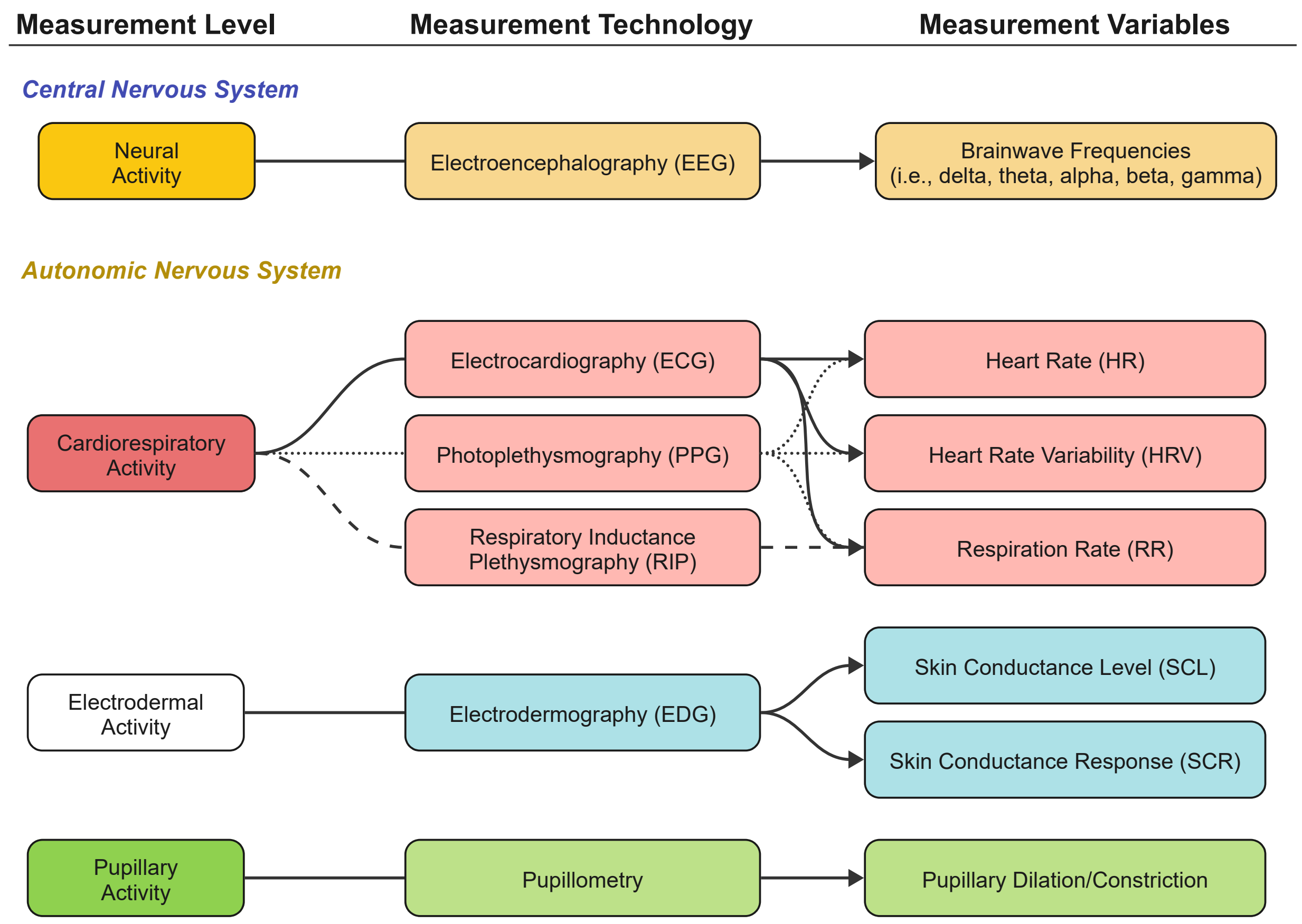

**Figure 1. Biosensing technologies and their measurements in research and applications in affective regulation and stress management.** While electrodermography is the measurement of skin conductance, "electrodermal activity" (EDA) and "galvanic skin response" (GSR) are more commonly used in the literature.

Research highlights music's affective regulation and thus therapeutic potential for stress management. Listening to music has been shown to decrease sympathetic nervous system activity.[15–17] This stress-relieving effect is attributed to two primary mechanisms. First, as a distraction from stressors, music can redirect attention,[18] and second, by triggering the release of dopamine in the reward system, music can induce relaxation.[19,20] These findings have spurred further investigations into stress reduction through music therapy (MT), music medicine (MM), and related interventions[21,22]. For instance, a meta-study[22] investigating the application of MT for treating stress and anxiety revealed various noteworthy outcomes, including psychological and physiological effects, distinctions between individual and group therapy settings, implementation of treatment protocols, and specific tempo and beat selections of music. However, both MT and MM are typically only administered by certified music therapists or healthcare providers. More recently, advances in sensing and computing have enabled the creation of systems that integrate musical elements, including tempo, rhythm, and tone, with biosensing technologies to support real-time affective regulation and stress management in any setting. Unlike MT and MM, these systems function without ongoing professional involvement, using physiological signals to personalize and modulate the musical experience. While promising, integrating music and biosensing for self-directed affective regulation and stress management remains an understudied area in current research. Thus, this review focuses on emerging systems that combine these elements to support scalable and adaptive affective regulation and stress management solutions.

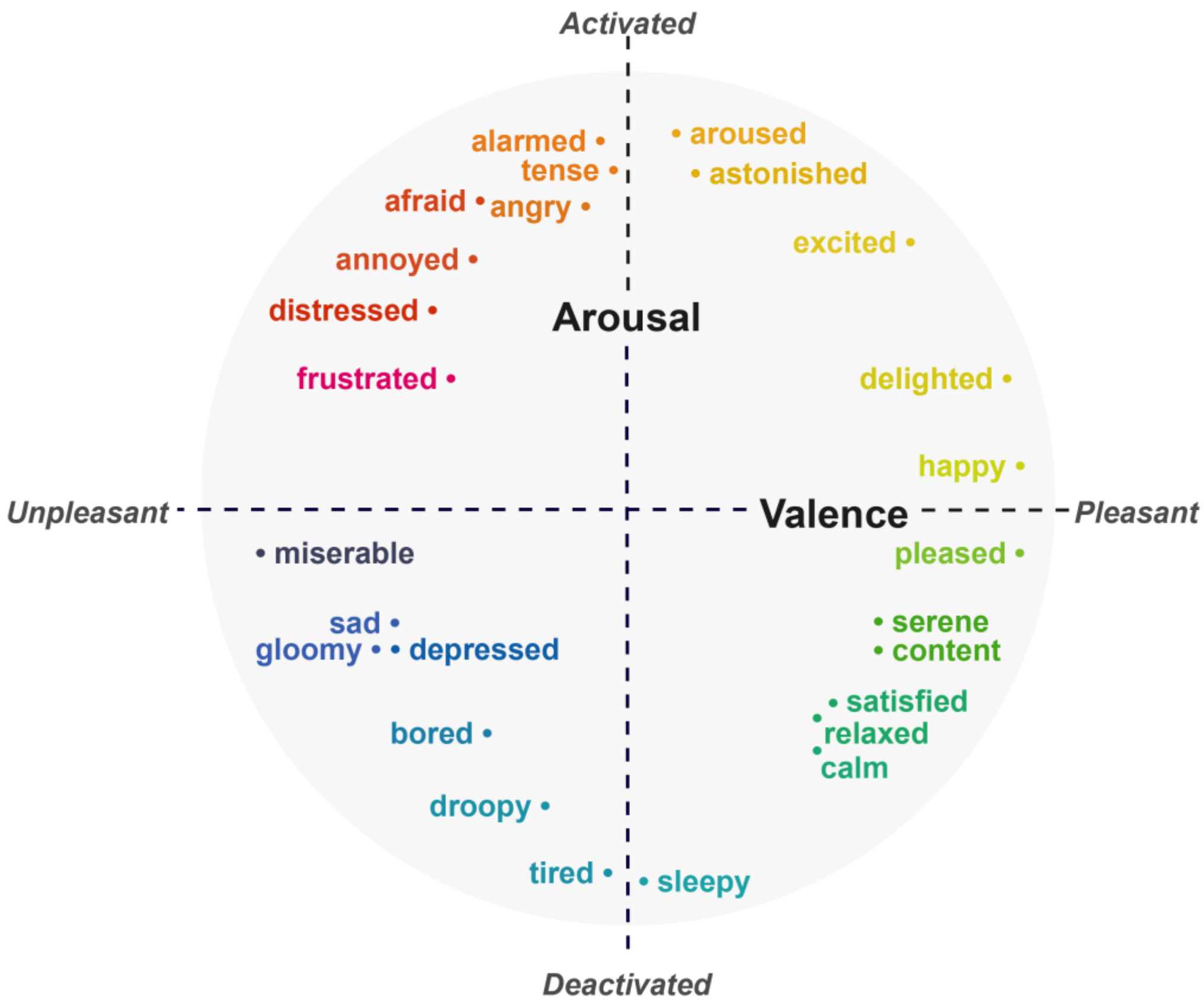

**Figure 2. Russell's (1980) circumplex model of affect.**

Computational approaches, such as machine learning (ML) and music information retrieval (MIR), play a pivotal role in the evolution of music-based affective regulation and stress management systems, driving innovations in music recommendation algorithms and more sophisticated and accurate emotion recognition models.[23,24] In MIR, researchers harness music properties by extracting and classifying elements such as tempo, rhythm, key, and melody to develop systems and methods that enhance how we interact with, recommend, and understand musical content. Together, ML and MIR techniques are pushing music emotion recognition (MER) to recognize and understand the emotional content of music. Music emotion recognition models leverage techniques such as feature extraction, wherein relevant musical features are identified, and classification algorithms, like support vector machines (SVMs) or neural networks (NNs), are used to categorize affective states. These models attempt to learn to generalize and predict emotional content in new music by training on labeled datasets that associate musical features with specific emotions. Simultaneously, ML techniques are used to infer users' affective states from multimodal data sources, such as biosensing or self-report measures. By combining information about music's emotional content and the user's affective state, these methods are spurring research in adaptive recommendation systems that customize music choices for entertainment and mental health purposes.[25,26] Finally, music recommender systems (MRS) that adapt preselected music to users' affective states are being explored as a means to regulate emotion and alleviate stress through music.

Given these rapidly growing research areas, this scoping and mapping review aims to summarize and synthesize the diverse landscape of systems and applications leveraging music and biosensing data for affective regulation and stress management. We draw from diverse sources, including systematic reviews and empirical research studies published in peer-reviewed journals and conference proceedings across the fields of psychology and computer sciences, and present a comprehensive overview of contemporary solution development and evaluation. These include systems that harness biosensing technology, further subcategorized according to the biosensing modality and interface type (e.g., mobile, ubiquitous) leveraged. Ultimately, this review aims to clarify how music and biosensing technologies intersect in the context of affective regulation and stress management, highlighting conceptual, methodological, and technical considerations to inform future system development and evaluation.

## 2. METHODOLOGY

This scoping and mapping review examines contemporary technologies and interfaces that utilize music for affective regulation and stress management. Unlike systematic reviews, this approach involves a more inductive and exploratory search strategy to explore a variety of studies[27] on digital tools and systems for music-based interventions. We provide quantitative and qualitative summaries for key categories of analysis, including biosensing modalities, types of music (e.g., prerecorded or generated), computational models (e.g., machine learning, deep learning), and biofeedback mechanisms. Within each category, we highlight a selection of representative studies to illustrate key trends, methodologies, and findings. Additionally, we categorize research studies and their respective systems according to other metadata variables, including study purpose, participant sample, and interface type.

### 2.1. Search Strategy

Our search strategy involved a purposive and iterative approach. We conducted initial searches across multiple academic research databases, including Google Scholar, PubMed, IEEE Xplore, and ACM Digital Library, using the following search terms, with truncation applied where appropriate: *music*, *technolog**, *interface**, *system*, *app**, *stress*, *anxi**, *emotion**, *affect**, *relax**. We screened the initial search results by title and filtered them as systematic reviews or empirical research studies. We also used a snowballing technique, reviewing references from the articles we identified to discover additional relevant literature.

Given increased advancements in biosensing technology in the last decade, we limit studies and articles in this review to those published from 2013 to the present. Inclusion criteria involved empirical research papers focusing on music-based technologies and interfaces that use biosensors and their applications for affective regulation and stress management. Exclusion criteria encompassed studies not published in English, non-peer-reviewed publications, and studies unrelated to music-based affective regulation and stress management systems. Such unrelated systems included those using auditory stimuli that are not strictly considered or manipulated concerning musical elements (i.e., rhythm, tempo, melodies, tones), such as white noise. We also excluded systems developed with biosensing and affect classification or modulation capabilities for applications in areas other than mental health, stress management, or mood improvement (e.g., entertainment, artistic creation).

### 2.2. Data Extraction and Synthesis

We systematically collected relevant information from the selected studies using a standardized data extraction form, including (1) publication year, (2) type of study conducted (i.e., technical feasibility, technology development, evaluation), (3) study purpose, (4) participant sample, and (5) study findings. This process aims to provide an overview of the current research landscape on music-based technologies and systems for affective regulation and stress management.

Data synthesis involved organizing and categorizing extracted information to identify relationships between music-based technologies and interfaces for affective regulation and stress management based on technologies, biosensing modality, computational model(s), music, and interfaces (i.e., desktop, mobile, wearable, ubiquitous) used. These categories were chosen to support the aim of uncovering emerging themes and identifying research gaps in developing and evaluating digital music-based affective regulation and stress management tools.

## 3. RESULTS

Results of this review demonstrate the diverse landscape of contemporary music-based biosensing technologies and interfaces for affective regulation and stress management. A total of $k$ = 28 studies describing systems evaluated across $n$ = 646 user participants were identified. We present the distribution of features across all systems reviewed in Figure 3. Among them, 53.6% use cardiorespiratory sensors: 21.4% electrocardiography (ECG), 10.7% photoplethysmography (PPG), 25.0% respiratory inductance plethysmography (RIP). Additionally, 21.4% use EDA sensors, and 32.1% use EEG technology. Of these, 60.7% use wearable systems (e.g., Zephyr BioHarness, Empatica E4). While desktop-based interfaces accounted for 71.4% of the total systems, mobile interfaces constituted 17.9%, ubiquitous types made up 21.4%, and immersive (i.e., virtual reality) interfaces comprised only 3.6%. In addition, 39.3% of the systems had a biofeedback mechanism, with most focusing on respiratory-based biofeedback. Further, advanced models or AI techniques were explicitly used in 28.6% of systems to detect affective states and 17.9% to predict music choices. Finally, 60.7% of the identified systems used prerecorded musical stimuli, including user- or experimenter-selected material, and 35.7% relied on generated music (i.e., music created by automated music composition systems or manipulated by the experimenters).

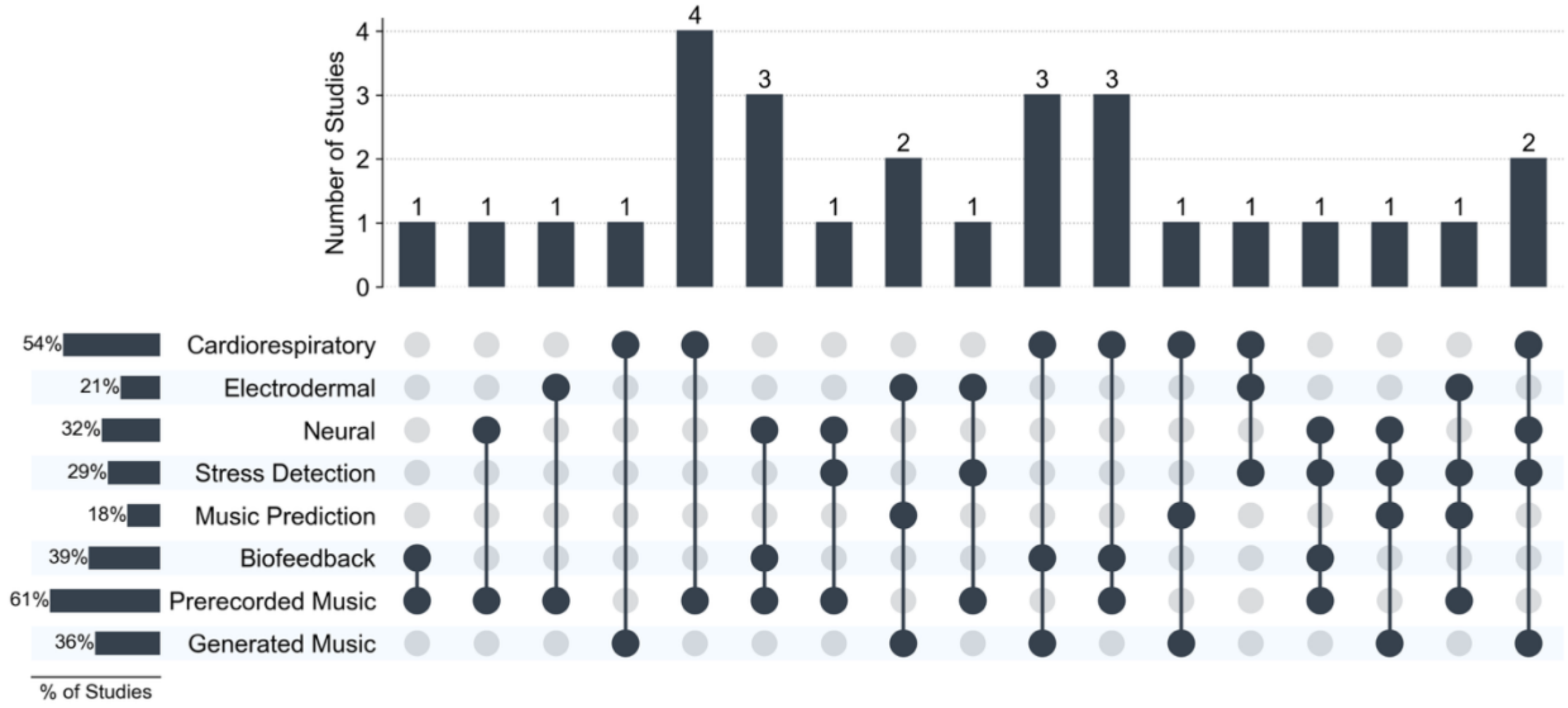

**Figure 3. Distribution of features across all *k* = 28 biosensing systems for music-based affective regulation, stress detection, and stress management.** "Stress Detection" and "Music Prediction" encompass studies using machine learning, deep learning, statistical, or other advanced computational models to classify affective states from biosensing data and/or predict music choices.

The reviewed studies suggest that most music-based systems developed for affective regulation or stress management show promise in inducing and modulating affective states. Overall, 53.6% of the systems, evaluated on $n$ = 317 user participants, demonstrate the ability to modulate users' affective states, as validated through statistically significant changes in self-report or physiological measures of stress, including heart rate (HR), heart rate variability (HRV), EDA, respiration rate (RR), and alpha and theta brainwaves. The remainder of the systems were assessed for their technical feasibility, such as using or creating various acoustic stimuli and detecting user affective states.

In the following subsections, we summarize the development and evaluation of digital music-based affective regulation and stress management tools, organized into four primary categories: physiological sensing, varieties of music, stress detection and music prediction models, and biofeedback systems. We subcategorize physiological sensing into cardiorespiratory, electrodermal, neural, and multimodal sensing and explore prerecorded and generated music from various genres for affective regulation and stress management. Stress detection and music prediction are discussed regarding specific computational models employed, including ML, deep learning (DL), and advanced statistical techniques. Finally, we discuss biofeedback systems, focusing on open-loop architectures that deliver pre-programmed feedback independent of user-specific biosensing inputs, and closed-loop architectures that dynamically adapt feedback in response to real-time biosensing data. A comprehensive summary of these studies is provided in Table 1.

### 3.1. Physiological Sensing

Below, we examine physiological sensing in music-based affective regulation and stress management systems, focusing on cardiorespiratory, electrodermal, and neural activity. We also discuss multimodal approaches that integrate multiple sensing modalities to monitor and respond to affective and stress-related physiological changes.

#### *3.1.1. Cardiorespiratory Sensing*

A meta-analysis by Kim and colleagues[28] demonstrates that cardiorespiratory activities are intrinsically linked to the stress response. Stressors can trigger the release of stress hormones, such as cortisol and adrenaline, leading to increased heart rate, blood pressure, sweating, and breathing rate. These responses are all involved in the adaptive mechanisms of the Autonomic Nervous System (ANS) when coping with perceived threats or challenges. Cardiorespiratory measures used to detect or corroborate the physiological stress response by the systems under review include HR, HRV, RR, and blood volume pulse (BVP). They are commonly recorded using ECG, PPG, and RIP.

Technologies integrating ECG, PPG, and RIP encompass stationary systems and ambulatory devices tailored for various medical and personal healthcare applications. These devices range from medical-grade monitors used in clinical settings, such as Holter and wired event monitors, to consumer-grade wearables, such as smartwatches and chest belts. Using such biosensing technologies, researchers interested in music-based affective regulation and stress management can develop more complex systems designed with algorithms that identify specific patterns of cardiorespiratory dynamics in the presence of a stressor. Studies have shown that low HRV conveys a monotonic HR and is associated with impaired regulatory and homeostatic ANS, reducing the body's ability to rapidly cope with internal and external stressors.[28] In this vein, elevations in HR and RR values, along with reduced HRV, can be used in these systems to prompt affective regulation or stress-reduction intervention delivery.

Over the last decade, evidence of developing and testing such systems that exclusively integrate cardiorespiratory sensing is present in $k$ = 13 studies involving $n$ = 243 participants. These include eight feasibility studies, three development studies, and two evaluation studies. Among the 13 systems employing HR or breathing, four exclusively use ECG,[29–32] two use only PPG,[33,34] five use RIP alone,[35–39] one integrates both ECG and RIP sensors,[40] and one uses breathing data with an unspecified biosensing technology.[41] All ECG-integrated systems collected data with wired electrodes (e.g., Biopac MP150) or wireless devices (e.g., Actiheart by CamNTech, Zephyr BioHarness).[30–32,40] The only exception was a system that relies on electromechanical film embedded in chair seats.[29] PPG- and RIP-integrated systems also leverage HR and breathing data collected with wireless chest belts[33,34-40] (e.g., Zephyr BioHarness, Vernier Go Direct) as well as finger-worn PPG sensors[32] (e.g., PureLight pulse oximeter). Most studies that created systems combining ECG and RIP through a chest belt opted for the well-validated Zephyr BioHarness.[42]

### *3.1.2. Electrodermal Sensing*

EDA or GSR refers to the skin's electrical conductance, which varies with changes in sweat gland activity driven by the ANS.[43] EDA can be used as a physiological indicator of affect and stress, reflecting sympathetic nervous system arousal. Among other physiological and contextual measures, EDA is used in stress detection algorithms to help identify changes in patterns that may signal the onset or escalation of physiological arousal.[44] Further, EDA is a common metric for validating arousal ratings when developing and evaluating affective regulation, stress detection, and stress management systems. Across all studies that met our inclusion criteria, only $k$ = 5 (involving $n$ = 196 user participants) music-based affective regulation and stress management systems exclusively collected EDA measures. Notably, EDA was not used directly in four systems for real-time stress detection or modulation. Instead, it was used as an outcome measure to validate the systems' abilities to detect or induce affective states.[45–48] One system, by Van der Zwaag and colleagues[47], incorporated EDA to inform real-time music selection based on users' arousal levels (see Section 3.3.1 for a detailed discussion). Two studies also focused on the feasibility and evaluation of music generators for prospective stress management applications.[45,46] EDA devices featured in these studies included the E4 (Empatica), Nexus-10 (MindMedia), BrainAmp GSR sensor (BrainProducts), and the Shimmer3 GSR+ Unit (Shimmer Sensing, Inc.).

An example study from our review corpus, Daly and colleagues[46] proposed and evaluated an affect-driven music generator for brain-computer music interfaces (BCMIs) to modulate users' affect to a target state. The ability of the system to generate novel musical stimuli corresponding to 9 possible affective states was validated by EDA recordings from $n$ = 20 listeners using the BrainAmp GSR sensor. Participants' music-induced and self-reported affect ratings were collected via FEELTRACE[49] and the Self-Assessment Manikin (SAM)[50] and then analyzed with EDA peak amplitudes and affective states targeted by the generator. The study found that, relative to baseline, EDA peak amplitude increased with self-reported arousal and decreased with higher, or more positive, self-reported valence, with statistically significant differences confirmed via bootstrapping ($p < .01$). Additionally, EDA peak amplitude significantly covaried with the targeted stress level of the music generation system. Together, findings of this evaluation demonstrate that the music generator induces a range of targeted affective states in its listeners and that EDA is a reliable biosignal for assessing and validating stress modulation in BCMIs.

Other exclusively EDA-based systems—affective music players and MER systems—with and without adaptive music manipulation, were also assessed for their affect modulation capabilities using EDA.[45,46,48,51] For instance, Bartolomé-Tomás and colleagues[48] evaluated a MER system designed to detect changes in physiological arousal levels in $n$ = 40 older adults in Spain using a variety of musical stimuli and EDA recordings. The musical stimuli consisted of prerecorded pieces custom-composed in styles representing various genres, including rock/jazz, Cuban, Spanish folklore, and Flamenco, and the EDA data were collected using the wrist-worn Empatica E4. Their objective was to investigate how familiarity with these musical genres influences affective responses, providing insights that could guide future intervention systems to foster emotional self-regulation in older adults. The study revealed significant temporal, morphological, statistical, and frequency differences across the genres. Among them, Flamenco and Spanish folklore music produced the highest number of statistically significant parameters associated with emotional arousal. These results highlight how the interplay of cultural background and musical familiarity may influence physiological responses to music, an important factor that could be explored further in future studies.

#### *3.1.3. Neural Sensing*

The complex neural activity related to affective regulation and stress involves interactions in brain regions such as the limbic system, hypothalamus, amygdala, and prefrontal cortex, orchestrating the stress response by balancing the ANS and the neuroendocrine system.[52] By interpreting neural signatures in such regions, brain-computer interfaces (BCIs) provide an approach to measuring and monitoring physiological levels, facilitating potential applications in adaptive brain-based affective regulation and stress management interventions. BCIs are computational systems that interpret brain signals, typically from EEG, and convert them into commands for output devices to perform tasks.[53] Thus, BCIs provide a direct link between the brain and a computer or other external device, allowing individuals to control machines using brain activity.[54] BCI systems generally comprise three components: (1) *data acquisition and preprocessing* to capture and clean brain signals, (2) *feature extraction* to identify relevant patterns in the signals, and (3) *classification* to decode features for controlling external devices.

Applications of BCI systems have been explored with music and other auditory stimuli to identify and mediate an individual's affective state. Such affective brain-computer music interfaces (aBCMIs) detect neural correlates of a user's current affective state and attempt to modulate it by generating or selecting calming or relaxing music.[55] Some researchers argue that aBCMIs offer an advantage over traditional music therapy by directly monitoring users' affective states through physiological indices, which may provide more robust and objective measures of emotion compared to self-reports or a music therapist's subjective appraisal.[56]

The past decade has witnessed a surge in publications on aBCMIs, simultaneously contributing to direct-to-consumer home applications of aBCMI technology.[55] The Mico system, a 2013 conceptual wearable device, offers personalized music choices through headphones and an iPhone application.[57] By analyzing brainwaves, the sensor in the headphones categorizes users into "neural groups," selecting music from a database that matches the identified neural pattern. Imec's EEG headset aims to measure and influence affect, learn users' musical preferences, and create real-time music to align with emotional states.[58] Neurosity's Crown™, a portable EEG device, is marketed as a productivity booster that detects brain waves and plays focus-enhancing music from Spotify.[59]

EEG features leveraged in music-based affective regulation and stress management systems include power spectral density across different frequency bands and alpha and beta wave activity. Alpha and beta waves, particularly in the 12 to 32 Hz range, have been identified as key indicators of physiological stress levels, with increased alpha-wave activity in the 8 to 12 Hz range commonly linked to relaxation and enhanced cognitive function.[60] While increasing the number of EEG channels improves stress detection accuracy, a growing focus has been on selecting the most effective channels to balance accuracy with practicality. Studies have shown that stress detection can still be effective with fewer channels, using configurations ranging from eight to even a single channel at the frontal site.[61,62] Combined with feature extraction and ML techniques, these optimized channel selections have achieved up to 81.6% accuracy rates in one study.[62]

Within the scope of this review, we identified $k$ = 9 neural sensing studies involving $n$ = 175 participants, of which 8 focus on developing or evaluating aBCMIs with a range of EEG devices, including wireless, ambulatory, and stationary systems with varying channel configurations.[63,64,66–71] Notably, most studies employing neural sensing did not specify the brands or models of EEG devices used. However, two systems used the 14-channel wireless Emotiv EPOC EEG headset,[64,65] while one used the 6-channel g.SAHARA Hybrid EEG system.[70] The earliest feasibility study using an aBCMI found in our timeframe was published in 2013. Uma and Sridhar[69] assessed the feasibility of a stationary EEG-based BCI system to recognize and manage stress levels based on the circumplex model of affect.[4,5] Three categories of musical stimuli (i.e., 'soft/melody,' 'devotional,' and 'rock/fast beat') were presented to participants while EEG waveforms were recorded using a 64-channel system. Relying on visual inspection to differentiate neural activity during exposure to music genres, the authors concluded that alpha, beta, and theta rhythms in frontal regions could be reliably differentiated across music categories and used for future classifier development. However, no ground truth data were obtained on participants' affective states or stress levels.

Another system by Tiraboschi et al.[70] incorporated the 6-channel g.SAHARA Hybrid EEG system (g.tec medical engineering GmbH) for real-time affective state classification and music generation with neural sensing. For EEG feature extraction, they computed the logarithmic root mean square of band-powers—theta, slow alpha, alpha, beta, and gamma. In addition, they evaluated different EEG channel configurations for real-time affective state classification in a BCI setup. They also compared the performance of systems using a minimal number of channels, testing both broader and more focused frequency bands to determine the most effective configuration for classifying affective states. Three setups were tailored to different needs: one optimized for robust features, another for minimal hardware, and a final one combining both approaches for valence classification.

#### *3.1.4. Multimodal Sensing*

By integrating peripheral physiological signals such as HR and EDA, along with central signals like brain waves, multimodal biosensing provides a more comprehensive approach to understanding physiological responses than relying on a single modality. Over the past decade, several studies have leveraged multimodal data to evaluate its potential to enhance affective regulation, stress detection, and stress management, with recent research fusing biosignal features illustrating improved accuracy and robustness[72–77]. For example, Kalimeri and Saitis[75] developed a multimodal framework using EEG and EDA signals to assess the emotional and cognitive experiences of blind and visually impaired individuals in unfamiliar indoor environments, achieving an automatic classification of environments with a 79.3% area under the receiver operating characteristic curve (AUROC) and providing insights into the biomarkers related to stress and cognitive load in varying situational contexts. Kim et al.[76] developed a shuffled ECA-Net deep neural network model that combines multiple biosignals, including ECGs, respiratory waveforms, and electrogastrograms, achieving 91.6% accuracy, 91.7% sensitivity, 91.6% specificity, 91.4% $F_1$ score, and 96.4% AUROC for stress detection, demonstrating the effectiveness of multimodal sensor fusion in improving psychological stress detection. While many of these studies highlight the growing potential of multimodal sensor systems for affective regulation and stress management, only $k$ = 2 studies involving $n$ = 52 user participants incorporated a multimodal approach in a music-based stress detection or management system.

Following their earlier work[46], Daly et al.[71] developed and evaluated an aBCMI to detect and modulate a user's current affective state with adaptive music choices. Their aBCMI system includes four separate processes: (1) multimodal physiological data acquisition, (2) affective state detection, (3) affective trajectory identification, and (4) music generation according to the affective trajectory. Multimodal data was collected with a 32-electrode EEG system, finger-worn EDA sensors, single-lead ECG electrodes on the wrists, and a respiration chest belt—all connected to the BrainAmp ExG amplifier (Brain Products, Germany). Their system captured a comprehensive range of physiological signals by combining EEG, EDA, ECG, and respiration sensor data, enabling a potentially more nuanced understanding of affective dynamics than a single biosensing modality could achieve. Specifically, the system employed a participant-specific automatic feature selection process to classify affective states, tailoring the selection of features to those that best represent each individual's emotional states. The process resulted in EEG features, predominantly from theta, alpha, and beta frequency bands in the right hemisphere, being selected 81.0% of the time. In comparison, physiological features were selected 19.0% of the time.

Ayata, Yaslan, and Kamasak[77] proposed an MRS based on user affect detected multimodally with physiological data from $n$ = 32 participants in the DEAP dataset.[85] The system uses an ML algorithm to predict affect from PPG and GSR data. PPG and GSR sensor data are inputted, segmented, and then processed in feature extraction to yield mean, maximum, minimum, or variance statistics. Their evaluation concluded that combining features from multiple modalities improved detection accuracies to 72.1% for arousal and 71.1% for valence, demonstrating the potential of multimodal affective data for music recommendation engines.

## 3.2. Varieties of Music for Stress Management

Auditory stimuli used by music-based affective regulation and stress management systems vary from prerecorded to generated music or music manipulated in different ways (e.g., volume modulation, channel reduction, or noise addition[35–37]) to provide variations in auditory cues. Prerecorded music relies on curated tracks with affective regulation or stress-relieving properties, as indicated by changes in physiological arousal or self-reported stress levels. In contrast, generated music uses algorithms to create compositions tailored to an individual's physiological state, emotional feedback, or personal preferences in real time.

### *3.2.1. Prerecorded Music*

Prerecorded music, commonly used in music-based affective regulation and stress management systems, relies on curated tracks with known affective and structural properties that have been found to promote relaxation and reduce stress. Several music genres, including classical (i.e., baroque, classical, and romantic periods), ambient soundscapes, instrumental tracks, and nature sounds, have been shown to modulate physiological and self-report measures of stress.[78,79] Among these genres, researchers have identified tempo and instrumentation as key musical properties contributing to these effects. Slower tempos in the 60 to 80 bpm range, such as in meditative music, are frequently linked to reductions in HR and increased relaxation.[80,81] Interestingly, instrumental music without vocals may be more effective for lowering physiological arousal levels than music with lyrics, as lyrics may be more distracting or stimulating.[82] However, this effect may vary culturally, and some researchers propose that lyrics can reduce physiological arousal by offering comforting messages that promote relaxation.[83]

We identified $k = 17$ systems tested with $n = 384$ participants that used prerecorded music selections. Among these, 11 featured multiple songs across different genres, 4 used a single genre (i.e., ambient/meditation, Western classical, or electronic),[31,40,51,65] and 3 studies used custom compositions.[38,41,48] Additionally, a study by Van der Zwaag et al.[47] evaluated their system's affective detection and modulation using 36 songs selected from participants' music libraries, based on their ratings of 200 randomly chosen tracks.

### *3.2.2. Generated Music*

Generated music, created with advanced AI algorithms, may offer significant methodological advantages over prerecorded tracks. Unlike prerecorded music, which relies on pre-existing compositions with fixed affective and structural properties, generated music can be tailored to meet specific experimental or therapeutic needs. Providing standardized stimuli ensures consistency across participants and sessions, reducing potential confounds caused by musical elements such as tempo, melody, and harmony variations. Additionally, AI-based music generation allows researchers to systematically customize individual music variables, such as rhythm, pitch, or timbre, enabling the study of their isolated effects on physiological regulation and other outcomes.

Music generators use various algorithms and methods to manipulate musical properties and their interactions with targeted affective states across different levels of abstraction (e.g., from low-level features like pitch and rhythm to higher-order structures like harmony). These constitute predominantly data-driven methods, including neural network architectures and Hidden Markov models, alongside rule-based, optimization, and hybrid approaches combining multiple strategies.[84] We identified $k = 10$ systems tested with $n = 230$ participants using generated music in our review.[34–37,45,46,48,64,70,71]

For example, in Daly et al.'s system,[46] a 16-channel feed-forward artificial neural network was trained on 12 bars of polyphonic piano music in C major at 120 bpm to transform seed structures (i.e., predefined musical patterns that serve as a foundation for variation). Musical elements—tempo, mode, pitch, timbre, and amplitude envelope—were mapped to different affective states via a Cartesian grid comprising valence and arousal, analogous to the circumplex model. For instance, a point with coordinates (1, 1), indicating low valence and low arousal, guided the generator to produce music with a slow tempo, minor chords, soft timbre, an amplitude envelope with considerable legato, and a narrow pitch range. This approach exemplifies how affective space can be operationalized via affect-to-music mapping and allows for controlled generation of emotional content in music.

Williams et al.'s[45] generative music system used a Hidden Markov Model (HMM) to generate music by learning probabilistic state transitions from source material and creating new musical permutations with controlled feature constraints. Specifically, the second-order HMM employed a musical feature matrix to model the likelihood of transitioning to a particular state based on current and preceding states. This approach enabled the system to create music following patterns learned from the training data while allowing variations in the generated sequences. The generative process enables discrete control over five musical parameters: pitch, rhythm, timbre, harmony, and tempo. EDA was used to measure physiological changes associated with the system's generated music, designed to align with specific points in a two-dimensional arousal-valence emotional space, ranging from 'low mindfulness' (i.e., high arousal, low valence) to 'high mindfulness' (i.e., low arousal, high valence). Overall, their evaluation provides insight into arousal responses evoked by generated music. However, whether mindfulness, an affective state involving cognitive and attentional control that was not directly assessed in the study, is fully captured by these indices remains an open question.

## 3.3. Stress Detection and Music Prediction Models

Various computational models have been applied to interpret physiological data underlying affective states and stress and to inform music generation or selection strategies. These include traditional statistical approaches and more advanced artificial intelligence techniques such as DL and generative adversarial networks (GANs). Researchers have developed and evaluated systems that explicitly integrate such models in a total of $k$ = 9 studies involving $n$ = 212 participants.[32,46,48,64,65,67,70,71,77]

### *3.3.1. Machine Learning and Statistical Models*

Traditional ML and statistical models typically rely on features extracted from biosignals, such as HRV, EDA, or EEG frequency bands that serve as peripheral physiological indicators of stress in the user. Statistical models like probability density functions (PDFs) and HMMs are used to analyze and model these physiological patterns, estimating probabilities or separations between the user's affective states. ML algorithms, such as support vector machines (SVMs), decision trees, and k-nearest neighbors (kNN), classify users' affective states related to stress. Such models rely on relatively low-dimensional, interpretable features and offer high-weight predictive performance while requiring minimal computational resources. Among the studies reviewed in this section, $k$ = 6 involving $n$ = 148 participants explicitly described using traditional ML algorithms or statistical models to predict affective states or music choice selections.[46,48,64,70,71,77]

In a multimodal aBCMI presented by Daly et al.,[46,71] the authors employed a shallow artificial neural network (ANN) to generate new music and an SVM to classify users' affective states, leveraging EEG frequency features alongside ECG, EDA, BVP, and RR measures. An ANN is a model consisting of interconnected layers of nodes (or "neurons") that process data by learning patterns and relationships within the input, in this case, the musical features. An SVM, on the other hand, is a supervised ML algorithm that identifies the optimal boundary (or hyperplane) to classify physiological data points into distinct affective categories by maximizing the distance between the boundary and nearest data points. In their first study[46], physiological data were recorded during music exposure to evaluate the generator's effectiveness in inducing affective responses. The authors evaluated how well the music generator's nine target affective states aligned with $n$ = 20 participants' subjective experiences, finding moderate correlations between participants' self-reported valence ($r$ = 0.60 $p$ < .01) and arousal ($r$ = 0.55, $p$ < .01) ratings. In a second study,[71] the authors used multimodal physiological features as input to train and test an SVM classifier to identify $n$ = 8 users' current affective states in real time. These classifications were passed to a case-based reasoning system that guided the music generator along predefined trajectories within the same arousal-valence space to modulate participants' affective states. Evaluation results revealed that the SVM could detect users' current affective states across three distinct classes with accuracies up to 65% (3 class, $p$ < .01), allowing the aBCMI to modulate users' affective states significantly above chance level ($p$ < .05).

Similarly, Bartolomé-Tomás and colleagues[48] evaluated different classifiers' abilities to detect physiological arousal from $n$ = 40 older adults' EDA during exposure to varying genres of music. The authors assessed multiple classifiers: decision trees, ensemble trees, logistic regression, linear discriminant analysis (LDA), Naïve Bayes, kNN, and SVM. Among these, SVM achieved the highest classification accuracy, reaching 87% for Flamenco and 83.1% for Spanish folklore music, while kNN showed competitive results with accuracies exceeding 80% for the same genres. While these classifiers are well established, the novelty of this work lies in its high classification performance for culturally specific music genres. The findings suggest that genre familiarity may enhance physiological responsiveness and classification accuracy in older adults and emphasize the importance of incorporating culturally and demographically tailored stimuli when developing music-based affect-aware systems.

More recently, aBCMIs have used LDA to classify affective states from EEG. In these systems, LDA is used to find the optimal linear combinations of EEG features to maximize the separation between different affective state classes (e.g., high versus low arousal or positive versus negative valence). Ehrlich et al.[64] implemented a hybrid approach, combining LDA and a rule-based probabilistic algorithm to classify two affective states (i.e., 'happy' and 'sad') from EEG features and to generate adaptive musical feedback tailored to the detected affective states, respectively. In later work, Tiraboschi et al.[70] evaluated the performance of LDA, a Naïve Bayes classifier, and an SVM to classify affective states using EEG features, with a focus on assessing whether a reduced number of EEG channels could support effective binary classification of valence and arousal. The classifiers were trained on features extracted from the DEAP dataset,[85] including power spectral densities from EEG signals collected during interactions of $n$ = 32 users with the system. The Naïve Bayes classifier, a probabilistic model based on Bayes' theorem, was used to estimate the likelihood of binary affective classes (i.e., positive/negative valence and low/high arousal) derived from normalized valence and arousal scores in the DEAP dataset. Among the models tested, LDA performed best in valence ($F_1$ = 0.61) and arousal ($F_1$ = 0.56) classification across various electrode configurations. While these results demonstrate the feasibility of EEG-based classification using a reduced sensor setup, the modest $F_1$ scores suggest limited discriminative power, which may reflect the absence of complementary autonomic signals such as HRV or EDA to improve affective state classification.

Two other systems exclusively employed statistical models to predict music choices. For example, in Van der Zwaag et al.'s affective music player,[47] probabilistic models were used to analyze EDA, predict users' affective states, and recommend music tracks. During their training phase, participants provided self-reported mood ratings, which were used to assess how different songs influenced skin conductance level (SCL). The system then used PDFs to model the likelihood of a song increasing, decreasing, or having a neutral effect on SCL, based on the AUROC for different ranges of delta SCL residuals (i.e., the differences in SCLs between successive songs, corrected for baseline SCL and the influence of the preceding song). In their test phase, the system used these probabilities to select tracks expected to induce an 'energized' state (increased SCL) or a 'calm' state (decreased SCL). From their evaluation on $n$ = 10 participants, the authors found that SCL in the 'up' condition (aimed at increasing arousal) was consistently higher than in the 'down' condition (aimed at decreasing arousal) from the first song onward ($F(2,14) = 6.39$, $p = .006$, $\eta^2 = 0.48$). Skin temperature was also higher in the 'up' condition compared to the 'down' condition from the fifth song onward ($F(2,14) = 6.96$, $p = .004$, $\eta^2 = .50$).

### *3.3.2. Deep Learning and Advanced Models*

Deep learning models, including convolutional neural networks (CNNs), recurrent neural networks (RNNs), and GANs, were explored in $k$ = 3 studies involving $n$ = 64 participants for processing biosensing data in physiological arousal prediction and affective music generation.[32,65,67] Compared to traditional ML models, DL approaches can learn complex, nonlinear patterns directly from raw biosignals, thus removing the need for manual feature engineering. CNNs analyze spatial features in signals like EEG and ECG, applying convolutional filters to identify stress-related patterns such as specific frequency band activity or HRV. RNNs model temporal dependencies in biosignals, enabling the classification of affective states based on changes over time. GANs are used for music generation, where the generator creates compositions conditioned on affective states detected from physiological features, and the discriminator evaluates whether the generated music aligns with the desired emotional "tone." Although DL can model more nuanced relationships in data, they typically require larger labeled datasets, greater compute resources, and offer less interpretability than traditional ML models. While DL has been widely applied to general stress detection from biosignals, few studies have explored its integration into music-based affective systems.

Kimmatkar and Babu[65] evaluated the performance of multiple classifiers in detecting participant-reported affective states (i.e., angry, calm, happy, or sad) from EEG data, three of which were neural networks (i.e., CNN, DNN, RNN), and one a kNN classifier. Participants' self-selected stimuli (i.e., thoughts, audio, or video) were used to induce one of the four targeted affective states while their EEG signals were recorded. A total of 24 EEG features were extracted with Chirplet transform from 14-channel EEG data collected from $n$ = 22 participants, with band power identified as a prominent feature. Their evaluation revealed that, while classification accuracy was modest overall (≤ 50%), kNN consistently outperformed the other classifiers across emotional categories. Although classification accuracy was the only reported performance metric, additional measures such as AUROC or $F_1$ score could have provided more insight into classifier behavior. Further, the authors' modest accuracy levels likely reflect a combination of factors, including the relatively small dataset, variability in self-selected emotional stimuli, and the inherent noise and ambiguity in EEG-based emotion labeling.

Idrobo-Ávila and colleagues[32] evaluated the feasibility of developing a biofeedback system based on GANs that could generate and alter sequences of harmonic musical intervals (HMIs), or chords, to elicit target HRV responses. The authors used two GANs (i.e., 'GAN-1' and 'GAN-2'), each formed by a generator and a discriminator. Each discriminator was trained to classify whether a given input sequence was a real or synthetic HMI or HRV data sequence generated by its respective generator. GAN-1's discriminator was trained with human-created HMIs and HMIs generated by its respective generator. It achieved a mean accuracy of 53% for human-created and 52% for generated data, indicating high similarity between data sources. Similarly, GAN-2's discriminator was trained with human HRV data and HRV data from audio data by its respective generator. The mean discrimination accuracy was 56% for real and 51% for generated data, suggesting good performance in generating new HRV data. While accuracy was reported as a proxy for discriminator performance, the calculation method was not clearly described. Additional evaluation metrics (e.g., loss values, AUROC) could help contextualize the GANs' performance, as accuracy values near 50% can be counterintuitive to interpret, particularly in generative settings where such values may reflect stronger generator performance. Although these initial results are promising at the model level, due to study constraints, the system could not be piloted and evaluated on human participants to assess whether HRV data could be modulated using generated HMIs. However, these findings contribute to a potential working model to implement generated music in a biofeedback-based stress management system.

### 3.4. Biofeedback Systems

Biofeedback-based stress management systems externalize an individual's internal physiological state, allowing users to monitor HR, HRV, or RR changes in real-time.[86] Most biofeedback systems use closed-loop architectures, in which real-time feedback is provided to the user based on their physiological data. Open-loop systems do not provide immediate feedback to the user but may record data for later analysis or intervention planning.

Biofeedback-based stress management systems enable users to learn to consciously regulate their physiological stress responses, such as incorporating controlled breathing techniques to facilitate parasympathetic nervous system activation.[87] Among the 28 systems in our review, $k$ = 11 involving $n$ = 176 participants incorporated a biofeedback mechanism,[32,34–41,46,66] of which 7 operate on respiratory-based biofeedback. These systems use auditory feedback—music and other sound cues—in addition to other modalities (e.g., visual, haptic) to prompt mindful or slow breathing.[35–41] While most systems employed a closed-loop architecture, Marentakis et al.[37] implemented an open-loop architecture, arguing that it may be less vulnerable to specific challenges associated with data collection, such as interference with user activities and data privacy concerns. Therefore, they evaluated three types of generated auditory feedback stimuli for guided breathing. The auditory stimuli consisted of (1) a synthesized pseudo-breath sound, (2) a musical sequence of notes rising and falling in pitch, and (3) a combination of both the synthetic breath and musical sequence. Ten adult volunteers were recruited to evaluate the ability of each auditory feedback type to guide their breathing along two target respiration rates—one slow and one fast—while wearing a respiration belt. Participants underwent multiple testing phases, including regular and paced breathing sections, with each phase featuring different counterbalanced types of feedback (i.e., breath, music, compound) and respiration rates. Results showed that all three types of feedback effectively guided participants to match the target breathing rate, with more significant deviations observed during fast breathing. Music feedback resulted in a more significant average deviation from the target breathing rate than breath feedback. However, compared to breath feedback alone, compound feedback demonstrated significantly more minor errors and longer durations close to the target breathing rate, particularly under conditions of fast respiration rates.

Although open-loop implementations address user convenience and data privacy challenges, they lack the personalization offered by closed-loop systems. With closed-loop systems, feedback can be dynamically adapted to users' real-time biosensing input. For example, Zepf et al.[40] developed a closed-loop biofeedback system to promote calm breathing in $n$ = 12 vehicle drivers through haptic and acoustic feedback. The system monitors the user's breathing patterns and delivers real-time rhythmic cues through three feedback modes: acoustic, haptic, and combination. Their study demonstrates that acoustic and mixed feedback significantly reduced participants' breathing rates compared to baseline (both $p$ = .03) without impairing focus during a simulated driving task. However, no statistically significant effects were observed for HR ($p$ = .97), HRV ($p$ = .10), or subjective ratings of stress, though participants rated the acoustic condition slightly lower in stress than in baseline (2.3 compared to 2.8 on a 7-point scale).

While many biofeedback systems target respiration to guide parasympathetic activation, other closed-loop systems use non-respiratory signals such as EEG and HRV. Yu et al.[34] implemented a closed-loop biofeedback system called Unwind, which dynamically modulates meditation music and nature sounds based on $n$ = 40 users' real-time HRV. The system continuously tracks users' HRV to assess arousal levels and adjusts the auditory feedback—blending calming nature sounds like flowing water or birdsong with meditation music—to modulate physiological arousal levels and promote relaxation and stress reduction. In another study, Ehrlich et al.[64] piloted a closed-loop aBCMI for emotion mediation with $n$ = 5 participants. In this system, user EEG patterns calibrated during an initial listening phase are continuously analyzed and mapped onto a valence-arousal space to generate adaptive music. Users in the study were able to intentionally modulate their affective states through the feedback music, demonstrating that EEG-informed musical biofeedback could serve as a regulatory tool.

## 4. DISCUSSION

The literature reviewed herein highlights how the integration of biosensing and music in interactive systems supports personalized and adaptive approaches to affective regulation and stress management by dynamically adjusting musical parameters in response to users' physiological signals. While growing research in this area has spurred the development of direct-to-consumer applications and devices that promote affective regulation and detect and reduce stress with auditory stimuli, many of these systems still lack research-grade results. This limitation highlights the need for more rigorous validation and standardization in the field, as existing research employs diverse methodologies of varying quality, leading to a broad range of results depending on their participants, datasets, experimental stimuli, recording protocols, affect and stress measures, and computational models. In light of these factors, we take an inductive approach to exploring the diverse array of contemporary biosensing systems that utilize music to regulate affect and reduce stress, while providing insights into future research directions and areas of improvement.

The results of our review highlight the diversity and evolution of music-based biosensing technologies and interfaces with potential applications in stress management. Twenty-eight systems with feasibility, development, and evaluation studies involving 646 participants across studies were identified. The systems reviewed encompass various biosensing modalities, with cardiorespiratory sensing being the most prevalent, followed by neural and electrodermal sensing. Various interfaces were employed across the systems, including desktop-based interfaces, mobile applications, wearable devices, ubiquitous systems, and incipient frameworks designed for immersive platforms. Desktop interfaces, particularly those integrating cardiorespiratory sensors, were predominant.

Within each biosensing modality, distinct trends in system development and evaluation emerged. Cardiorespiratory sensing systems predominantly use ECG, PPG, and RIP technologies, often integrated into stationary and ambulatory devices. These systems frequently incorporate various forms of prerecorded and generated musical stimuli, the latter of which may present a methodological advantage in providing standardized stimuli for stress interventions. Additionally, biofeedback mechanisms, particularly those focusing on respiratory-based biofeedback with closed-loop architectures, are prevalent in systems focused on mindful breathing. The integration of auditory cues in biofeedback systems aimed at affective regulation and stress reduction through breathwork is gaining traction. However, additional investigation is warranted to determine optimal characteristics of auditory cues to support personalization and increase the effectiveness of music-based biofeedback interventions.

Electrodermal activity appears to be a key biosignal in music-based affective regulation and stress reduction tools. Particularly in feasibility and evaluation studies, EDA often serves as a reliable measure of autonomic arousal against which researchers have validated outputs from affect and stress detection or music prediction models. However, fewer recent systems utilize music and EDA for affective regulation or stress management, likely due to the lack of reliable, non-intrusive wrist-worn EDA devices. While finger-worn sensors are more reliable for capturing EDA, they may not be as practical for continuous, real-world use due to their intrusiveness. This could limit the integration of EDA into more convenient and user-friendly wearable systems for everyday stress management.

Neural sensing systems leverage EEG data to detect and modulate users' affective states through personalized music interventions. Among all biosensing systems reviewed, neural sensing systems dominated as a mobile form factor, highlighting ongoing efforts towards increasing accessibility to real-time, on-the-go interventions.[15] Recent advances in wearable EEG technology are making neural sensing more accessible through compact, less obtrusive headsets that are more suitable for everyday use. However, variability in access to software development kits and evidence of signal validation across brands poses barriers for researchers to leverage consumer-oriented devices. EEG-based systems also continue to face technical challenges, including high cost, user discomfort, and susceptibility to noise.

Despite the diverse landscape of systems reviewed and their promise, researchers should consider the following cross-cutting challenges relating to system design and evaluation, including technical considerations around sensing and portability and the conceptual distinction between affective regulation and stress management.

Sensor reliability and stability, particularly in real-world contexts, remain persistent obstacles, and data artifact-handling remains a central and often unreported issue in physiological signal processing. As the field places increasing emphasis on biosensing-integrated interventions, the need to standardize physiological signal quality control and preprocessing procedures becomes critical. Standardization in these areas is essential for enabling robust, transparent, and reproducible inferences from autonomic nervous system data across varied application contexts.[88] Additionally, desktop-based systems have provided a valuable foundation for integrating biosensing and music-based affect regulation and stress management. However, their limited portability and scalability pose challenges for in-the-wild deployment, where mobility and contextual responsiveness are increasingly important.

While many biosensing systems in our review explicitly focus on affective regulation with appropriate validation measures, suggesting potential for future stress management applications, several others were positioned as stress management tools without using validated stress measures. This reflects a broader challenge in how stress is conceptualized and measured. Although the circumplex model of affect provides a useful framework for understanding arousal and valence, researchers should carefully consider whether these constructs align with how stress is framed and validated in their systems. In particular, distinguishing between physiological reduction (e.g., an objective decrease in heart rate or skin conductance) and subjective relaxation (e.g., a self-reported feeling of calm) requires thoughtful selection of stress measures, whether investigator-determined or participant-reported. This distinction ensures that measurement approaches yield accurate, meaningful insights and support the design of effective systems.

Trends in computational modeling appear to vary across systems. While cardiorespiratory sensing systems rely primarily on rule-based models for inferring affect from physiological data and predicting music choices, EDA and neural sensing systems extensively apply ML and DL techniques. Traditional ML classifiers like SVMs, kNN, and LDA were commonly employed due to their relatively low computational complexity and interpretability. Models in these systems typically rely on engineered features and achieve moderate classification accuracies, with some systems reporting up to 87% accuracy for specific genres or affective targets. More recent studies have integrated DL architectures, including CNNs, RNNs, and GANs that can learn complex nonlinear patterns from biosignal data, but often do not report comprehensive performance metrics, limiting interpretability and comparability. Among these, GANs have emerged as a novel approach to generate adaptive music, where generators create compositions tailored to affective states and discriminators ensure that the generated music aligns with affective targets. Although GAN systems remain in early development and have yet to be widely tested with human participants, preliminary findings show their potential to generate biosignal-responsive music sequences.

While ML and DL methods for affective regulation, stress detection, and stress management have advanced, challenges remain in interpreting ANS responses and integrating these models into biofeedback systems for practical use. A meta-analysis of 202 studies examining ANS reactivity during induced emotions in non-clinical adults demonstrates increased effect sizes for most ANS variables across emotion categories but no clear differentiation between categories.[89] These findings suggest that ANS responses are context-specific and highly variable and that researchers should exercise caution when developing and training ML models to predict stress. This variability also limits the reliability of biofeedback systems that rely on non-adaptive models or single-sensor input. Data fusion from multiple contextual sources and sensing modalities, which has shown promising results in enhancing emotion recognition accuracy [73], may help capture the variability of users' affective experiences.

Finally, most existing systems rely on prerecorded music selections, with multiple pieces across multiple genres selected for their relaxation-inducing properties. However, using various pieces when designing and evaluating music-based interventions can pose methodological challenges due to inconsistencies in affective and structural characteristics across pieces. In addition, findings from interventions using culturally specific genres—such as Flamenco or Spanish folklore—underscore that musical preferences are culturally bound,[48,83] adding another layer of complexity to study design and measurement selection to account for cultural variance. Such potential confounds can be more rigorously controlled in feasibility studies by testing with a single genre or piece. Given these challenges, the growing use of generated music, especially through generative AI technologies, offers a promising avenue for future work.[90] Music-based emotion regulation and stress management systems can infuse such technologies to create more personalized music that is adaptive to a user's various affective contexts.

## 5. CONCLUSION AND FUTURE DIRECTIONS

This scoping and mapping review underscores the wealth of biosensing systems utilizing music for affective regulation and stress management that have emerged in the last decade. Over half of the systems reviewed demonstrate the capability to induce physiological or self-reported changes in affective state, indicating potential in affective regulation and stress management system development. However, while the studies reviewed offer researchers and engineers valuable insight into refining existing systems to support mental health with biosensing and music, notable gaps in the research exist.

First, considerable variability exists in evaluation metrics, system design choices, and theoretical framing, particularly concerning the distinction between affective regulation and stress reduction. Many systems are positioned within the context of stress management but validate outcomes using affective models without stress-specific measures. This conceptual and methodological ambiguity limits comparability across studies and may obscure how different systems actually influence stress-related outcomes. Future work on stress management intervention design should aim for greater precision in articulating which psychological or physiological constructs are being measured and how those constructs are being validated.

A second, closely related issue concerns reporting practices. Many studies lack sufficient detail about physiological data preprocessing, classifier parameters, or evaluation protocols. In particular, limited reporting of performance metrics (e.g., reliance on raw accuracy without AUROC, $F_1$ score, or effect sizes) restricts the interpretability of findings. As the field moves toward more adaptive biosensing systems for music-based affect regulation and stress management, transparent, rigorous, and standardized reporting practices will be essential to ensure reproducibility and enable comparable evaluation of system effectiveness.

Third, there is a need for more investigations integrating multimodal biosensing approaches with music-based affect detection and regulation. In general stress research studies, the combination of data collected from multiple modalities has been shown to improve the performance of stress detection models.[72–75] However, some researchers argue that instead of combining as many data sources as possible, selecting modalities for use in a stress detection framework should balance prediction accuracy and other crucial evaluation criteria, such as intrusiveness, user comfort, privacy, and scalability.[91] This suggests that future investigations should aim to identify the most accurate combination of modalities for music-based stress detection and consider their feasibility and appropriateness for real-world application.

Fourth, this domain can benefit from more studies identifying specific stress-reducing properties of a musical genre or piece. Many systems rely on genre labels (e.g., ambient, meditative) without systematically evaluating which musical features contribute to relaxation or arousal modulation. Music information retrieval techniques may allow researchers to extract specific music features from such genres for further testing and integration into stress management models using prerecorded or AI-generated music.

Finally, as systems increasingly rely on collecting and analyzing sensitive user health data, concerns about data privacy and protection become crucial. Some experts have argued for users' rights to mental privacy and integrity.[92,93] Ensuring the confidentiality and security of this data is essential to gaining and maintaining usage and trust. This will require that the individual user be in control of what is recorded, how the recordings are stored, and what is revealed and shared by the system about their mental health data and classification results.

## ABBREVIATIONS

| | |
|---|---|
| aBCMI | Affective brain-computer music interface |
| AI | Artificial intelligence |
| ANN | Artificial neural network |
| ANS | Autonomic Nervous System |
| AUROC | Area under the receiver operating characteristic curve |
| BCI | Brain-computer interface |
| BVP | Blood volume pulse |
| CNN | Convolutional neural network |
| DL | Deep learning |
| ECG | Electrocardiography |
| EDA | Electrodermal activity |
| EEG | Electroencephalography |
| GAN | Generative adversarial network |
| GSR | Galvanic skin response |
| HMI | Harmonic music interval |
| HMM | Hidden Markov Model |
| HR | Heart rate |
| HRV | Heart rate variability |
| kNN | k-nearest neighbors |
| LDA | Linear discriminant analysis |
| MER | Music emotion recognition |
| MIR | Music information retrieval |
| ML | Machine learning |
| MM | Music medicine |
| MRS | Music recommender system |
| MT | Music therapy |
| NN | Neural network |
| PDF | Probability density function |
| PPG | Photoplethysmography |
| RIP | Respiratory inductance plethysmography |
| RNN | Recurrent neural network |
| RR | Respiration rate |
| SCL | Skin conductance level |
| SVM | Support vector machine |

## REFERENCES

1. Chrousos, G. P. & Gold, P. W. The concepts of stress and stress system disorders: Overview of physical and behavioral homeostasis. *JAMA* **267**, 1244–1252 (1992).
2. Mariotti, A. The effects of chronic stress on health: New insights into the molecular mechanisms of brain–body communication. *Future Sci. OA* **1**, 1–6 (2015).
3. Seaward, B. L. *Managing Stress: Principles and Strategies for Health and Well-Being*. (Jones & Bartlett Learning, Burlington, MA, 2017).
4. Can, Y. S., Arnrich, B. & Ersoy, C. Stress detection in daily life scenarios using smart phones and wearable sensors: A survey. *J. Biomed. Inform.* **92**, 103139 (2019).
5. Luštrek, M., Lukan, J., Bolliger, L., Lauwerier, E. & Clays, E. Designing an intervention against occupational stress based on ubiquitous stress and context detection. In: *Adjunct Proceedings of the 2023 ACM International Joint Conference on Pervasive and Ubiquitous Computing & the 2023 ACM International Symposium on Wearable Computing* 606–610 (Association for Computing Machinery, New York, NY, 2023).
6. Alberdi, A., Aztiria, A. & Basarab, A. Towards an automatic early stress recognition system for office environments based on multimodal measurements: A review. *J. Biomed. Inform.* **59**, 49–75 (2016).
7. Hickey, B. A. *et al.* Smart devices and wearable technologies to detect and monitor mental health conditions and stress: A systematic review. *Sensors* **21**, 3461 (2021).
8. Yu, B., Funk, M., Hu, J., Wang, Q. & Feijs, L. Biofeedback for everyday stress management: A systematic review. *Front. ICT* **5**, 1–22 (2018).
9. Russell, J. A. A circumplex model of affect. *J. Pers. Soc. Psychol.* **39**, 1161–1178 (1980).
10. Posner, J., Russell, J. A. & Peterson, B. S. The circumplex model of affect: An integrative approach to affective neuroscience, cognitive development, and psychopathology. *Dev. Psychopathol.* **17**, 715–734 (2005).
11. Stanisławski, K. The coping circumplex model: An integrative model of the structure of coping with stress. *Front. Psychol.* **10**, 1–23 (2019).
12. Coulon, S. M., Monroe, C. M. & West, D. S. A systematic, multi-domain review of mobile smartphone apps for evidence-based stress management. *Am. J. Prev. Med.* **51**, 95–105 (2016).
13. Riches, S., Azevedo, L., Bird, L., Pisani, S. & Valmaggia, L. Virtual reality relaxation for the general population: A systematic review. *Soc. Psychiatry Psychiatr. Epidemiol.* **56**, 1707–1727 (2021).
14. Jin, S., Kim, B. & Han, K. "I don't know why I should use this app": Holistic analysis on user engagement challenges in mobile mental health. *Proc. 2025 CHI Conf. Hum. Factors Comput. Syst.* 1–23 (2025).
15. Bartlett, D. L. Physiological responses to music and sound stimuli. In: *Handbook of Music Psychology* (ed. Hodges, D. A.) 343–385 (IMR Press, San Antonio, TX, 1996).
16. Linnemann, A., Strahler, J. & Nater, U. M. Assessing the effects of music listening on psychobiological stress in daily life. *J. Vis. Exp.* 1–9 (2017) https://doi.org/10.3791/54920.
17. Halbert, J. D. *et al.* Low frequency music slows heart rate and decreases sympathetic activity. *Music Med.* **10**, 180–185 (2018).
18. Kiss, L. & Linnell, K. J. The effect of preferred background music on task-focus in sustained attention. *Psychol. Res.* **85**, 2313–2325 (2021).

19. Salimpoor, V. *et al.* The rewarding aspects of music listening involve the dopaminergic striatal reward systems of the brain: An investigation with [C11]Raclopride PET and fMRI. *NeuroImage* **47**, S160 (2009).
20. Belfi, A. M. & Loui, P. Musical anhedonia and rewards of music listening: Current advances and a proposed model. *Ann. N. Y. Acad. Sci.* **1464**, 99–114 (2020).
21. de Witte, M., Spruit, A., van Hooren, S., Moonen, X. & Stams, G.-J. Effects of music interventions on stress-related outcomes: A systematic review and two meta-analyses. *Health Psychol. Rev.* **14**, 294–324 (2020).
22. de Witte, M. *et al.* Music therapy for stress reduction: A systematic review and meta-analysis. *Health Psychol. Rev.* **16**, 134–159 (2022).
23. Shinde, A. S. *et al.* ML based speech emotion recognition framework for music therapy suggestion system. In: *2022 6th International Conference On Computing, Communication, Control And Automation (ICCUBEA)* 1–5 (IEEE, Pune, India, 2022). https://doi.org/10.1109/ICCUBEA54992.2022.10011091.
24. Chiang, W. C., Wang, J. S. & Hsu, Y. L. A music emotion recognition algorithm with hierarchical SVM based classifiers. In: *2014 International Symposium on Computer, Consumer and Control* 1249–1252 (2014). https://doi.org/10.1109/IS3C.2014.323.
25. He, J. Algorithm composition and emotion recognition based on machine learning. *Comput. Intell. Neurosci.* **2022**, 1–10 (2022).
26. Ceccato, C., Pruss, E., Vrins, A., Prinsen, J. & Alimardani, M. BrainiBeats: A dual brain-computer interface for musical composition using inter-brain synchrony and emotional valence. In: *Extended Abstracts of the 2023 CHI Conference on Human Factors in Computing Systems* 1–7 (Association for Computing Machinery, New York, NY, 2023). https://doi.org/10.1145/3544549.3585910.
27. Grant, M. J. & Booth, A. A typology of reviews: An analysis of 14 review types and associated methodologies. *Health Inf. Libr. J.* **26**, 91–108 (2009).
28. Kim, H.-G., Cheon, E.-J., Bai, D.-S., Lee, Y. H. & Koo, B.-H. Stress and heart rate variability: A meta-analysis and review of the literature. *Psychiatry Investig.* **15**, 235–245 (2018).
29. Liu, H., Hu, J. & Rauterberg, M. Follow your heart: Heart rate controlled music recommendation for low stress air travel. *Interact. Stud.* **16**, 303–339 (2015).
30. Zhu, Y., Wang, Y., Li, G. & Guo, X. Recognizing and releasing drivers' negative emotions by using music: Evidence from driver anger. In: *Adjunct Proceedings of the 8th International Conference on Automotive User Interfaces and Interactive Vehicular Applications* 173–178 (Association for Computing Machinery, New York, NY, 2016). https://doi.org/10.1145/3004323.3004344.
31. Zhu, B., Hedman, A., Feng, S., Li, H. & Osika, W. Designing, prototyping and evaluating digital mindfulness applications: A case study of mindful breathing for stress reduction. *J. Med. Internet Res.* **19**, e6955 (2017).
32. Idrobo-Ávila, E., Loaiza-Correa, H., Muñoz-Bolaños, F., van Noorden, L. & Vargas-Cañas, R. Development of a biofeedback system using harmonic musical intervals to control heart rate variability with a generative adversarial network. *Biomed. Signal Process. Control.* **71**, 103095 (2022).
33. Shin, I.-H. *et al.* Automatic stress-relieving music recommendation system based on photoplethysmography-derived heart rate variability analysis. *Annu. Int. Conf. IEEE Eng. Med. Biol. Soc.* **2014**, 6402–6405 (2014).

34. Yu, B., Funk, M., Hu, J. & Feijs, L. Unwind: A musical biofeedback for relaxation assistance. *Behav. Inf. Technol.* **37**, 800–814 (2018).
35. Harris, J., Vance, S., Fernandes, O., Parnandi, A. & Gutierrez-Osuna, R. Sonic respiration: Controlling respiration rate through auditory biofeedback. In: *CHI '14 Extended Abstracts on Human Factors in Computing Systems* 2383–2388 (Association for Computing Machinery, New York, NY, 2014). https://doi.org/10.1145/2559206.2581233.
36. Leslie, G., Ghandeharioun, A., Zhou, D. & Picard, R. W. Engineering music to slow breathing and invite relaxed physiology. In: *2019 8th International Conference on Affective Computing and Intelligent Interaction (ACII)* 1–7 (IEEE, Cambridge, UK, 2019). https://doi.org/10.1109/ACII.2019.8925531.
37. Marentakis, G., Borthakur, D., Batchelor, P., Andersen, J. P. & Grace, V. Using breath-like cues for guided breathing. In: *Extended Abstracts of the 2021 CHI Conference on Human Factors in Computing Systems* 1–7 (ACM, Yokohama Japan, 2021). https://doi.org/10.1145/3411763.3451796.
38. Sato, T. G., Ooishi, Y., Fujino, M. & Moriya, T. Device for controlling the phasic relationship between melodic sound and respiration and its effect on the change in respiration rate. *Behav. Inf. Technol.* **0**, 1–13 (2023).
39. Bhandari, R., Parnandi, A., Shipp, E., Ahmed, B. & Gutierrez-Osuna, R. Music-based respiratory biofeedback in visually-demanding tasks. In: *Proceedings of the International Conference on New Interfaces for Musical Expression* 78–82 (Baton Rouge, LA, 2015).
40. Zepf, S., Kao, P.-W., Krämer, J.-P. & Scholl, P. Breath-triggered haptic and acoustic guides to support effortless calm breathing. In: *2021 43rd Annual International Conference of the IEEE Engineering in Medicine & Biology Society (EMBC)* 1796–1800 (IEEE, Virtual Conference, 2021). https://doi.org/10.1109/EMBC46164.2021.9629766.
41. Shor, D., Ruitenburg, Y., Boere, W., Lomas, J. D. & Huisman, G. The Resonance Pod: Applying haptics in a multi-sensory experience to promote relaxation through breathing entrainment. In: *2021 IEEE World Haptics Conference (WHC)* 1143–1143 (IEEE, Montreal, QC, Canada, 2021). https://doi.org/10.1109/WHC49131.2021.9517165.
42. Nazari, G. *et al.* Psychometric properties of the Zephyr BioHarness device: A systematic review. *BMC Sports Sci. Med. Rehabil.* **10**, 1–8 (2018).
43. Boucsein, W. *Electrodermal Activity*. (Springer Science+Business Media, Berlin, Germany, 2012).
44. Liu, Y. & Du, S. Psychological stress level detection based on electrodermal activity. *Behav. Brain Res.* **341**, 50–53 (2018).
45. Williams, D. *et al.* AI and automatic music generation for mindfulness. In: *2019 AES International Conference on Immersive and Interactive Audio: Creating the Next Dimension of Sound Experience* (Curran Associates, Inc., York, UK, 2019).
46. Daly, I. *et al.* Towards human-computer music interaction: Evaluation of an affectively-driven music generator via galvanic skin response measures. In: *2015 7th Computer Science and Electronic Engineering Conference (CEEC)* 87–92 (IEEE, Colchester, UK, 2015). https://doi.org/10.1109/CEEC.2015.7332705.
47. van der Zwaag, M. D., Janssen, J. H. & Westerink, J. H. D. M. Directing physiology and mood through music: Validation of an affective music player. *IEEE Trans. Affect. Comput.* **4**, 57–68 (2013).

48. Bartolomé-Tomás, A., Sánchez-Reolid, R., Fernández-Sotos, A., Latorre, J. M. & Fernández-Caballero, A. Arousal detection in elderly people from electrodermal activity using musical stimuli. *Sensors* **20**, 4788 (2020).
49. Cowie, R. *et al.* 'FEELTRACE': An instrument for recording perceived emotion in real time. In: *Proceedings of the ISCA Workshop on Speech and Emotion* (Newcastle, Northern Ireland, UK, 2000).
50. Bradley, M. M. & Lang, P. J. Measuring emotion: The Self-Assessment Manikin and the semantic differential. *J. Behav. Ther. Exp. Psychiatry* **25**, 49–59 (1994).
51. Qin, Y., Zhang, H., Wang, Y., Mao, M. & Chen, F. 3D music impact on autonomic nervous system response and its therapeutic potential. In: *2020 IEEE Conference on Multimedia Information Processing and Retrieval (MIPR)* 364–369 (IEEE, Shenzhen, China, 2020). https://doi.org/10.1109/MIPR49039.2020.00080.
52. Buijs, R. M. & Van Eden, C. G. The integration of stress by the hypothalamus, amygdala and prefrontal cortex: Balance between the autonomic nervous system and the neuroendocrine system. In: *Progress in Brain Research* vol. 126 117–132 (Elsevier, 2000).
53. Shih, J. J., Krusienski, D. J. & Wolpaw, J. R. Brain-computer interfaces in medicine. *Mayo Clin Proc* **87**, 268–279 (2012).
54. Saha, S. *et al.* Progress in brain computer interface: challenges and opportunities. *Front. Syst. Neurosci.* **15**, 1–20 (2021).
55. Hildt, E. Affective brain-computer music interfaces—drivers and implications. *Front. Hum. Neurosci.* **15**, 1–4 (2021).
56. Williams, D. A. H. & Miranda, E. R. BCI for music making: Then, now, and next. In: *Brain–Computer Interfaces Handbook* (eds. Nam, C. S., Nijholt, A. & Lotte, F.) 193–206 (CRC Press, Boca Raton, FL, 2018).
57. NeuroWear. Mico: A smart headband for the mind. *NeuroWear* https://www.neurowear.com/mico.
58. Imec. Wearable EEG solutions. https://www.imec-int.com/drupal/sites/default/files/2019-01/EEG_Headset_digital.pdf (2019).
59. Neurosity. Crown: The brain-computer interface for productivity. *Neurosity* https://www.neurosity.co.
60. Asif, A., Majid, M. & Anwar, S. M. Human stress classification using EEG signals in response to music tracks. *Comput. Biol. Med.* **107**, 182–196 (2019).
61. Umar Saeed, S. M., Anwar, S. M., Majid, M., Awais, M. & Alnowami, M. Selection of neural oscillatory features for human stress classification with single channel EEG headset. *Biomed. Res. Int.* **2018**, 1049257 (2018).
62. Hag, A., Al-Shargie, F., Handayani, D. & Asadi, H. Mental stress classification based on selected electroencephalography channels using correlation coefficient of Hjorth parameters. *Brain Sci.* **13**, 1340 (2023).
63. Jayaraj, P. J., Ghazali, M. & Gaber, A. Relax app: Mobile brain-computer interface app to reduce stress among students. In: *Special Proceedings of 2021 Asian CHI Symposium* 92–96 (Virtual, 2021).
64. Ehrlich, S. K., Agres, K. R., Guan, C. & Cheng, G. A closed-loop, music-based brain-computer interface for emotion mediation. *PLoS One* **14**, 1–24 (2019).
65. Kimmatkar, N. V. & Babu, B. V. Novel approach for emotion detection and stabilizing mental state by using machine learning techniques. *Computers* **10**, 37 (2021).

66. Chen, H. M., Chen, S. Y., Jheng, T. J. & Chang, S. C. Design of a mobile brain-computer interface system with personalized emotional feedback. In: *Future Information Technology-II*, 87–95 (Springer Netherlands, 2015).
67. Sun, M. Study on antidepressant emotion regulation based on feedback analysis of music therapy with brain-computer interface. *Comput. Math. Methods Med.* 1–14 (2022).
68. Tiwari, A. & Tiwari, R. Design and implementation of a brain computer interface for stress management using LabVIEW. In: *2017 International Conference on Computer, Communications and Electronics (Comptelix)*, 152–157 (IEEE, 2017). https://doi.org/10.1109/COMPTELIX.2017.8003955.
69. Uma, M. & Sridhar, S. S. A feasibility study for developing an emotional control system through brain computer interface. In: *2013 International Conference on Human Computer Interactions (ICHCI)* 1–6 (IEEE, Chennai, India, 2013). https://doi.org/10.1109/ICHCI-IEEE.2013.6887801.
70. Tiraboschi, M., Avanzini, F. & Boccignone, G. Listen to your mind's (he)art: A system for affective music generation via brain-computer interface. In: *Proceedings of the 18th Sound and Music Computing Conference* (Zenodo, Virtual, 2021). https://doi.org/10.5281/zenodo.5044984.
71. Daly, I. *et al.* Affective brain–computer music interfacing. *J. Neural Eng.* **13**, 1–14 (2016).
72. Xefteris, V.-R. *et al.* A multimodal late fusion framework for physiological sensor and audio-signal-based stress detection: An experimental study and public dataset. *Electronics* **12**, 1–15 (2023).
73. Pinto, G. *et al.* Multimodal emotion evaluation: A physiological model for cost-effective emotion classification. *Sensors* **20**, 3510 (2020).
74. Lee, S., Lee, T., Yang, T., Yoon, C. & Kim, S.-P. Detection of drivers' anxiety invoked by driving situations using multimodal biosignals. *Processes* **8**, 155 (2020).
75. Kalimeri, K. & Saitis, C. Exploring multimodal biosignal features for stress detection during indoor mobility. In: *Proceedings of the 18th ACM International Conference on Multimodal Interaction* 53–60 (Association for Computing Machinery, New York, NY, 2016). https://doi.org/10.1145/2993148.2993159.
76. Kim, N., Lee, S., Kim, J., Choi, S. Y. & Park, S.-M. Shuffled ECA-Net for stress detection from multimodal wearable sensor data. *Comput. Biol. Med.* **183**, 109217 (2024).
77. Ayata, D., Yaslan, Y. & Kamasak, M. E. Emotion based music recommendation system using wearable physiological sensors. *IEEE Trans. Consum. Electron.* **64**, 196–203 (2018).
78. Burns, J. L. *et al.* The effects of different types of music on perceived and physiological measures of stress. *J. Music Ther.* **39**, 101–116 (2002).
79. Chennafi, M., Khan, M. A., Li, G., Lian, Y. & Wang, G. Study of music effect on mental stress relief based on heart rate variability. In: *2018 IEEE Asia Pacific Conference on Circuits and Systems (APCCAS)* 131–134 (IEEE, Chengdu, China, 2018). https://doi.org/10.1109/APCCAS.2018.8605674.
80. Hilz, M. J. *et al.* Music induces different cardiac autonomic arousal effects in young and older persons. *Auton. Neurosci.* **183**, 83–93 (2014).
81. Nomura, S., Yoshimura, K. & Kurosawa, Y. A pilot study on the effect of music-heart beat feedback system on human heart activity. *Journal of Medical Informatics & Technologies* **22**, 251–256 (2013).
82. Souza, A. S. & Leal Barbosa, L. C. Should we turn off the music? Music with lyrics interferes with cognitive tasks. *J. Cogn.* **6**, 24 (2023).

83. Good, M. *et al.* Cultural differences in music chosen for pain relief. *J. Holist. Nurs.* **18**, 245–260 (2000).
84. Dash, A. & Agres, K. AI-based affective music generation systems: A review of methods and challenges. *ACM Comput. Surv.* **56**, 287:1-287:34 (2024).
85. Koelstra, S. *et al.* DEAP: A database for emotion analysis using physiological signals. *IEEE Trans. Affect. Comput.* **3**, 18–31 (2012).
86. Schwartz, M. S. A new improved universally accepted official definition of biofeedback: Where did it come from? Why? Who did it? Who is it for? What's next? *Biofeedback* **38**, 88–90 (2010).
87. Fincham, G. W., Strauss, C., Montero-Marin, J. & Cavanagh, K. Effect of breathwork on stress and mental health: A meta-analysis of randomised-controlled trials. *Sci. Rep.* **13**, 432 (2023).
88. Dunn, J. *et al.* Building an open-source community to enhance autonomic nervous system signal analysis: DBDP-autonomic. *Front. Digit. Health* **6**, (2025).
89. Siegel, E. H. *et al.* Emotion fingerprints or emotion populations? A meta-analytic investigation of autonomic features of emotion categories. *Psychol. Bull.* **144**, 343–393 (2018).
90. Civit, M., Civit-Masot, J., Cuadrado, F. & Escalona, M. J. A systematic review of artificial intelligence-based music generation: Scope, applications, and future trends. *Expert Syst. Appl.* **209**, 118190 (2022).
91. Naegelin, M. *et al.* An interpretable machine learning approach to multimodal stress detection in a simulated office environment. *J. Biomed. Inform.* **139**, 104299 (2023).
92. Lavazza, A. Freedom of thought and mental integrity: The moral requirements for any neural prosthesis. *Front. Neurosci.* **12**, 82 (2018).
93. Ienca, M. & Andorno, R. Towards new human rights in the age of neuroscience and neurotechnology. *Life Sci. Soc. Policy* **13**, 5 (2017).

**Table 1. Chronologically Ordered Music-Based Affective Regulation and Stress Management Systems with Biosensing Integration (*k* = 28 studies; *n* = 646 participants)**

| Study | Study Type | Purpose | Type of Music | Biosensing Modality | Interface | Computational Model | Sample | Outcome |
|---|---|---|---|---|---|---|---|---|
| Uma and Sridhar (2013) | Feasibility | Assess the feasibility of developing a BCI system to recognize and control affective states using EEG frequencies and preselected music. | Prerecorded; Various | EEG | Desktop | Not described | *N* = 4; 2 F, 2 M; ages not reported | Alpha, beta, and theta rhythms in the frontal regions could be reliably differentiated during exposure to different music categories. |
| Van der Zwaag et al. (2013) | Feasibility | Validate whether an affective music player could induce changes in the direction of two "energized" or "calm" affective states | Prerecorded; Various | EDA | Desktop | Probabilistic model | *N* = 10; 5 F, 5 M; mean age: 26.5 ± 3.5 yrs | Skin conductance and mood could be directed toward energized or calm states, which persisted for at least 30 minutes. |
| Harris et al. (2014) | Feasibility | Present and validate "Sonic Respiration," an auditory biofeedback system to slow breathing rate for stress management using two forms of acoustic manipulation. | Generated; "On the Line" by James May | RIP | Mobile; Wearable | Not described | *N* = 6; 4 F, 2 M; age range: 20–59 yrs | Both forms of acoustic manipulation (i.e., adding white noise, reducing channels in a multi-track song) are equally effective at slowing breathing. |
| Shin et al. (2014) | Evaluation | Evaluate a wearable, wireless PPG-based stress-relieving music recommendation system. | Prerecorded; Various songs | PPG | Desktop; Wearable | Time-frequency analysis (to com- pute sympatho-vagal balance index [SVI]) | *N* = 20; 12 F, 10 M; mean age: 17.6 ± 2.7 yrs | The system showed strong correlations between SVI changes and participants' physiological responses to different music pieces, enabling stress-relieving music pieces to be identified. |
| Bhandari et al. (2015) | Feasibility | Present and evaluate a music-based respiratory biofeedback intervention to regulate stress levels during a visually-demanding task. | Prerecorded; Various slow-tempo songs | RIP | Desktop; Mobile; Wearable | Not described | *N* = 20; 5 F, 23 M; age range: 23–35 yrs | When compared to two non-biofeedback conditions, music biofeedback led to lower arousal levels across RR, HRV, and EDA measures. |
| Chen et al. (2015) | Development | Propose the development of a mobile aBCMI aimed at delivering real-time personalized emotional feedback to users. | Prerecorded; Various | EEG | Mobile; Wearable | Proprietary algorithm; Threshold-based affect scoring | Not described | The system can collect training data during exposure to various multimedia and output real-time data to users on their mental states. |
| Daly et al. (2015) | Evaluation | Evaluate an affectively-driven music generator for use in a BCMI to induce intended affective states in users. | Generated; Various | EDA* | Desktop | Artificial neural network | *N* = 20; 9 F, 11 M; mean age: 22.0 ± 1.5 yrs | There were moderate correlations between the generator’s targeted affective states and self-report valence and arousal ratings, indicating that the generator can induce targeted emotions in listeners. |
| Liu, Hu, and Rauter-berg (2015) | Development | Present a heart rate-controlled in-flight music recommendation system for stress reduction during air travel. | Prerecorded; Various user-selected songs | ECG (via electro-mechanical film) | Ubiquitous | Context- and content-based filtering | *N* = 12; 6 F, 6 M; age range: 25–35 yrs | A simulated long-haul flight experiment revealed that passengers' stress can be reduced through listening to music playlists preselected for decreasing, increasing, or maintaining user HR. |
| Daly et al. (2016) | Evaluation | Develop and evaluate an aBCMI for modulating the affective states of its users. | Generated; Various | ECG; EEG; RIP | Desktop; Wearable | SVM | *N* = 8; 6 F, 2 M; age range: 20–23 yrs | The system can detect users' affective states with classification accuracies of up to 65% (3 class, $p < .01$) and modulate its user's affective states ($p < .05$). |
| Zhu et al. (2016) | Feasibility | Assess the efficacy of recognizing negative affect through HR data and whether tempo and personal familiarity with the music can reduce drivers' negative affect, and consequently improve driving performance. | Prerecorded; Various user-selected songs | ECG | Ubiquitous; Wearable | Fourier analysis | *N* = 30; 12 F, 18 M; mean age: 22.7 ± 1.4 yrs | In a simulated driving experiment, HR data could be used in the recognition of driver anger. Medium-tempo music led to faster alleviation of negative affect compared to fast-tempo music. |
| Tiwari and Tiwari (2017) | Development | Propose the development of a mobile aBCMI to prompt the user via text messaging to engage in relaxation methods with yoga or listening to preselected music. | Prerecorded; Various | EEG; EOG | Desktop; Mobile; Wearable | Stream processing algorithm | *N* = 50; demographics not reported | The system was able to detect user states of stressed, stressed and relaxed by music, and stressed and relaxed by yoga from 82% of participants. |
| Zhu et al. (2017) | Development | Develop and test the feasibility of a physical digital mindfulness prototype for stress reduction. | Prerecorded; Meditation music | ECG | Ubiquitous; Wearable | Time-frequency analysis | *N* = 25; 13 F, 12 M; age range: 23–60 yrs | The prototype, incorporating vapor, light, and sonification, was effective in promoting mindful breathing and reducing stress levels, as indicated by both subjective self-assessment and HRV measures. |
| Ayata, Yaslan, and Kamasak (2018) | Feasibility | Propose an emotion-based music recommendation framework that learns user emotions based on EDA and PPG data. | Not described | EDA; PPG | Desktop; Wearable | Decision tree; KNN; Random forest; SVM | *N* = 32; 16 F, 16 M; mean age: 24.9 ± 4.5 yrs | Feature fusion with a multimodal sensor dataset increased the SVM classifier's accuracy rate compared to single modality. |
| Yu et al. (2018) | Evaluation | Evaluate "Unwind," a musical interface for a HRV biofeedback system that facilitates breathing regulation and relaxation. | Generated; Sedative music with nature sounds | PPG | Desktop; Wearable | Not described | *N* = 40; 22 F, 18 M; age range: 20–30 yrs | There was a significant interaction effect between music and biofeedback on the improvement of heart rate variability. |
| Williams et al. (2019) | Feasibility | Assess the feasibility of a generative music system to creating emotionally congruent music for applications in entertainment and mindfulness. | Generated | EDA* | Desktop; Wearable | Hidden Markov Model | *N* = 53; demographics not reported | The two types of music (i.e., tense–scary, calm–not scary) elicited emotional responses that matched participants' questionnaire descriptions with their EDA measures. |
| Ehrlich et al. (2019) | Evaluation | Develop and evaluate a BCI prototype that can feedback a user's affective state in a closed-loop interaction between EEG and musical stimuli. | Generated; Various | EEG | Desktop; Wearable (wireless Emotiv Epoc+) | LDA; Rule-based probabilistic model | Study 1: *N* = 11 4 F, 7 M; mean age: 26.9 ± 3.4 yrs<br><br>Study 2: *N* = 5; all male; mean age: 27.8 ± 5.0 yrs | In *Study 1*, there was a good match between users' perceptual ratings of affect and music generation settings, although there was high variance across subjects. In *Study 2*, participants were able to intentionally modulate the musical feedback by self-inducing emotions (e.g., recalling emotional memories). |
| Leslie et al. (2019) | Feasibility | Evaluate the feasibility of an interactive music system in influencing a user's breathing rate to induce a relaxation response across three interaction designs. | Generated; Ambient music with shifts in loudness | RIP | Desktop; Wearable | Rule-based; Breathing-based amplitude modulation | *N* = 19; 11 F, 8 M; age range: 19–55 yrs | The interactive music system effectively reduced breathing rates and physiological arousal, with the “personalized tempo” design having the largest effect. |
| Bartolomé-Tomás et al. (2020) | Feasibility | Assess the feasibility of detecting changes in arousal using musical stimuli and EDA measures of older individuals. | Prerecorded; Custom compositions in styles of four genres | EDA* | Desktop; Wearable | Time-frequency analysis; Logistic regression; LDA; Naïve Bayes; Decision trees; KNN; SVM | *N* = 40; 23 F, 17 M; mean age: 66.3 ± 5.9 yrs | Flamenco and Spanish Folklore music yielded the most number of significant EDA parameters. SVM and KNN showed the highest accuracies in arousal detection (> 80% for these genres). |
| Qin et al. (2020) | Feasibility | Evaluate the feasibility of using 3D music to modulate EDA responses in VR-based therapy for stress and anxiety. | Prerecorded; Electronic | EDA* | Immersive; Wearable | Not described | *N* = 73; 43 F, 30 M; age range: 12–66+ yrs | EDA can serve as an indicator of ANS activity and emotional arousal level, with 3D music significantly reducing EDA compared to other musical elements like tempo. |
| Jayaraj, Ghazali, and Ghaber (2021) | Development | Propose the design of a mobile aBCMI application to reduce stress among college students using human-computer interaction design principles. | Prerecorded; Solfeggio frequency; Binaural beats | EEG | Mobile | Not described | *N* = 11 (initial user needs survey); *N* = 6 (feasibility); *N* = 10 (usability); demographics not reported | Usability testing of the mobile BCI prototype revealed that the app showed good overall usability, with some inconsistencies noted. Most of the participants preferred the Solfeggio Frequency approach over binaural beats in reducing stress levels. |
| Kimmatkar and Babu (2021) | Feasibility | Detect emotional state by processing EEG signals and test the effect of meditation music therapy to stabilize mental state. | Prerecorded; Meditation music | EEG | Desktop; Wearable (wireless Emotiv Epoc+) | CNN; DNN; kNN; RNN | *N* = 22; 15 F, 7 M; mean age: 35.6 ± 17.0 yrs | The kNN classifier showed highest accuracy in classifying emotions. 75% of EEG signals from participants successfully transformed from the "annoying" state to the "relaxed" state. |
| Marentakis et al. (2021) | Feasibility | Compare three synthetic auditory feedback stimuli (i.e., breath, music, and compound) for guided breathing in an open-loop biofeedback system. | Generated; Various melodies and sounds | RIP | Desktop; Wearable | Not described | *N* = 10; 5 F, 5 M; mean age: 33.9 ± 12.7 yrs | Compound auditory feedback stimuli (i.e., synthetic breath and musical stimuli combined) show a stronger effect on breath entrainment to a target breathing rate. |
| Shor et al. (2021) | Feasibility | Explore the potential role of haptics as part of the “Resonance Pod,” an enclosed hanging chair using lights, music, and vibrations to combat stress through breathing entrainment. | Prerecorded; Custom composition for the system | Unknown sensor type; RR | Ubiquitous | Not described | *N* = 5; demographics not reported | Qualitative user feedback on four 3-minute breathing rhythm sequences suggests that the Resonance Pod creates a pleasant and calming multisensory breathing entrainment experience. |
| Tiraboschi, Avanzini, and Boccignone (2021) | Evaluation | Explore strategies for real-time music generation applications using biosensor data and evaluate the performance of supervised learning methods on classification of affective valence and arousal. | Generated; Various | EEG | Desktop | LDA; Naïve Bayes; SVM | *N* = 32; 16 F, 16 M; mean age: 24.9 ± 4.5 yrs | The pipeline can generate affectively driven music using EEG data. A reduced number of EEG channels can still be used for binary classification of affective valence and arousal. |
| Zepf et al. (2021) | Development | Present a closed-loop system that monitors breathing in real-time and provides rhythmical feedback (i.e., acoustic, haptic, and mixed) to support slow breathing and relaxation. | Prerecorded; Ambient music[a] | ECG; RIP | Ubiquitous; Wearable | Breathing-based feedback rate adaptation | *N* = 12; 5 F, 7 M; mean age: 31.3 ± 4.5 yrs | Acoustic and mixed feedback can slow breathing without affecting focus, suggesting that subtle rhythmic feedback can be an effective stimuli type in biofeedback systems. |
| Idrobo-Ávila et al. (2022) | Feasibility | Propose a HRV-based biofeedback system that can generate harmonic musical intervals to moderate HRV responses. | Generated; Harmonic music intervals (HMIs) | ECG | Desktop | GAN | *N* = 26; 9 F, 17 M; mean age: 25.3 ± 7.1 yrs | Using HRV data from human subjects, the GAN achieved comparable accuracy in generating HMIs to human-created HMIs, suggesting the potential use of HRV data to generate HMIs. |
| Sun (2022) | Evaluation | Propose and evaluate a feedback-based aBCMI for depression. | Prerecorded; Various | EEG | Desktop | CNN | *N* = 16 (4 controls, 8 depression, 4 feedback training); demographics not reported | Participants receiving neurofeedback training with the aBCMI showed lower self-reported depression ratings. |
| Sato et al. (2023) | Feasibility | Explore the potential of cyclic melodies to individualize the phasic relationship between sound and respiration (PRSR). | Western classical-style music | RIP | Desktop | Not described | *N* = 10; 8 F, 2 M; mean age: 40.6 ± 5.9 yrs | Respiration intervals can be changed by controlling the PRSR, suggesting that for biofeedback devices for daily use, the PRSR could be considered when melody is presented as a stimulus. |

*Note*. Technologies: aBCMI = affective brain-computer music interface; BCI = brain-computer interface. Biosensing modalities: ECG = electrocardiography; EEG = electroencephalography; EOG = electrooculography (eye blinks); HR = heart rate; HRV = heart rate variability; RIP = respiratory inductance plethysmography; RR = respiration rate. Computational models: LDA = Linear discriminant analysis; CNN = convolutional neural network; DNN = deep neural network; GAN = generative adversarial network; kNN = k-nearest neighbor; RNN = recurrent neural network; SVM = support vector machine.

[a] The selected ambient music can be found at: https://www.youtube.com/watch?v=n0svuurLibQ

* EDA was used as a validation signal during system evaluation and was not directly used by the system for real-time stress detection or modulation.